\documentclass[conference]{IEEEtran}
\IEEEoverridecommandlockouts
\usepackage{cite}
\usepackage{amsmath,amssymb,amsfonts}
\usepackage{algorithmic}
\usepackage{graphicx}
\usepackage{textcomp}
\usepackage{xcolor}
\usepackage{subcaption}
\usepackage{caption}
\DeclareCaptionJustification{justified}{\justifying}
\usepackage[raggedrightboxes]{ragged2e}
\usepackage{algorithm}
\usepackage{algorithmic}
\usepackage{float}
\usepackage{multirow}
\usepackage{makecell}
\usepackage{hyperref}

\def\BibTeX{{\rm B\kern-.05em{\sc i\kern-.025em b}\kern-.08em
    T\kern-.1667em\lower.7ex\hbox{E}\kern-.125emX}}

\newcommand\blfootnote[1]{%
  \begingroup
  \renewcommand\thefootnote{}\footnote{#1}%
  \addtocounter{footnote}{-1}%
  \endgroup
}

\begin{document}

\title{Exploring Autoregressive Vision Foundation Models for Image Compression 
}
\author{%
\IEEEauthorblockN{
Huu-Tai Phung
\quad  Yu-Hsiang Lin
\quad Yen-Kuan Ho
\quad Wen-Hsiao Peng
}
\IEEEauthorblockA{%
National Yang Ming Chiao Tung University, Taiwan
}
}


\maketitle

\IEEEpubidadjcol

\begin{abstract}
This work presents the first attempt to repurpose vision foundation models (VFMs) as image codecs, aiming to explore their generation capability for low-rate image compression. VFMs are widely employed in both conditional and unconditional generation scenarios across diverse downstream tasks, e.g., physical AI applications. Many VFMs employ an encoder-decoder architecture similar to that of end-to-end learned image codecs and learn an autoregressive (AR) model to perform next-token prediction. To enable compression, we repurpose the AR model in VFM for entropy coding the next token based on previously coded tokens. This approach deviates from early semantic compression efforts that rely solely on conditional generation for reconstructing input images. Extensive experiments and analysis are conducted to compare VFM-based codec to current SOTA codecs optimized for distortion or perceptual quality. Notably, certain pre-trained, general-purpose VFMs demonstrate superior perceptual quality at extremely low bitrates compared to specialized learned image codecs. This finding paves the way for a promising research direction that leverages VFMs for low-rate, semantically rich image compression.

\end{abstract}




\vspace{-0.5cm}
\blfootnote{This work is supported by National Science and Technology Council, Taiwan, under the Grant NSTC 113-2634-F-A49-007- and National Center for High-performance Computing, Taiwan.}

\section{Introduction}
\begin{figure*}[t]
    \centering
    \includegraphics[width=0.75\linewidth]{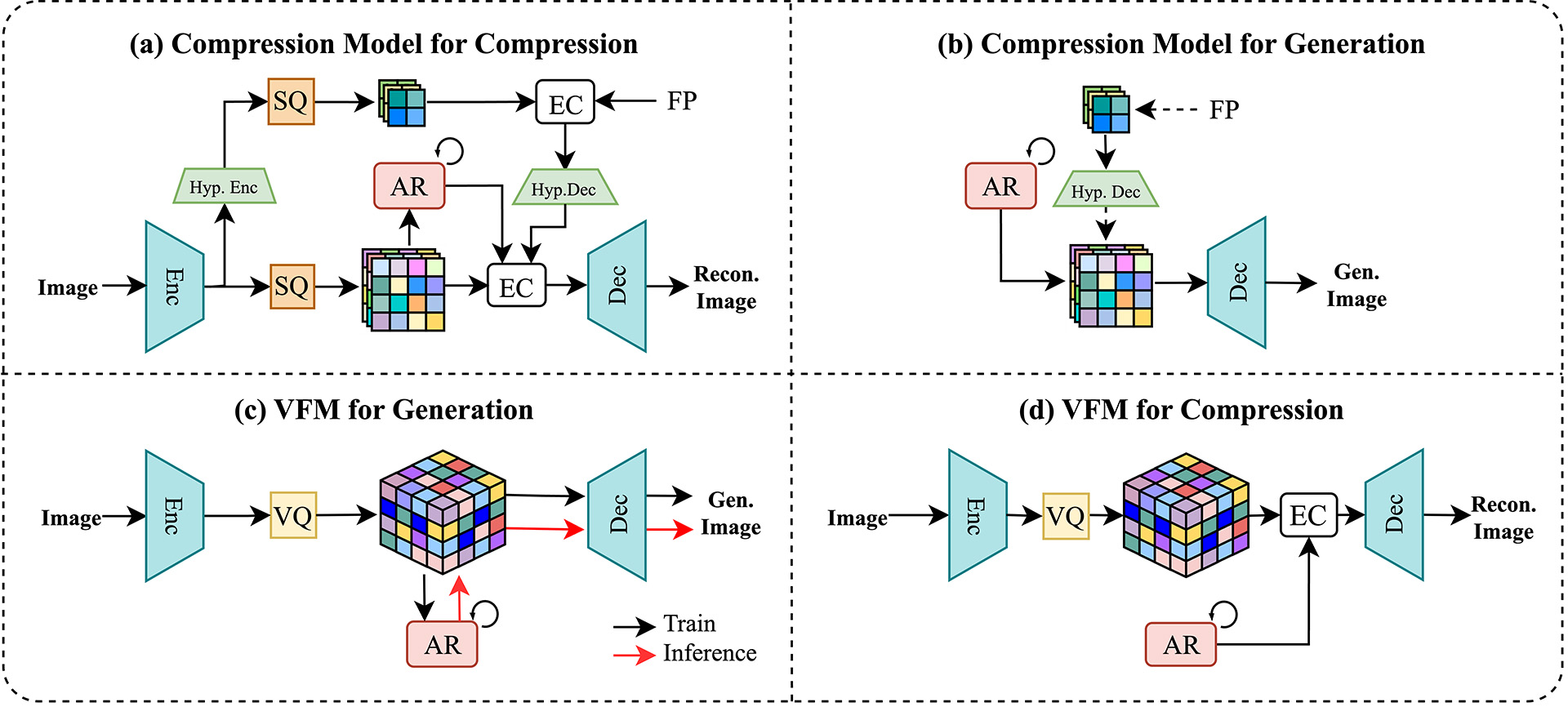}
    \caption{A comparative overview of the operational pipelines for compression and generation tasks. This figure illustrates how  learned image compression models and vision foundation models (VFMs) are adapted to both tasks. Abbreviations: SQ (scalar quantization), VQ (vector quantization), AR (autoregressive model), FP (factorized prior), EC (entropy coding).}
    \label{fig:teaser}
    \vspace{-1.5em}
\end{figure*}

Image compression and generation are two sides of the same coin. The seminal work by Ballé et al.~\cite{googleiclr18} highlights that training a rate-distortion-optimized learned image codec is effectively equivalent to learning a variational autoencoder (VAE). As illustrated in Fig. \ref{fig:teaser}(a), the compression process involves encoding an input image into its latent representations, quantizing these latents via scalar quantization (SQ), and modeling their distribution--often via an autoregressive (AR) model and a hyperprior--for entropy coding. These quantized latents are then decoded to reconstruct the input image. Notably, the decoder in such a compression model can be repurposed for unconditional image generation, as shown in Fig.~\ref{fig:teaser}(b). By sampling the hyperprior latents from the factorized prior (FP) distribution and the main image latents from the AR model along with the resulting hyperprior, and passing these generated latents through the main decoder, the system effectively transforms into a generative model.

Recent AR-based vision foundation models (VFMs) share a similar encode-decoder architecture to that of learned image codecs. As depicted in Fig.~\ref{fig:teaser}(c), a key distinction is that VFMs employ vector quantization (VQ) to transform image latents into discrete tokens, followed by training an AR model to perform next-token prediction. These VFMs can be deployed as world models in reinforcement learning settings, serving as environment simulators to facilitate agent training. Their primary objectives are to generate diverse and realistic images or videos. Toward this goal, they are often built on large models (e.g., with 12 billion network parameters~\cite{Cosmos}) or trained on extensive datasets (e.g. 20 million hours of video~\cite{Cosmos}). The exceptional generation quality of these large-scale VFMs stems from their highly accurate next-token prediction. We posit that this powerful predictive capability is not solely useful for generation; it is the hallmark of a potent statistical model that may find applications in image compression. This insight forms the cornerstone of our work: we investigate whether the VFM's AR model can be directly repurposed for entropy coding, a concept visualized in Fig. \ref{fig:teaser}(d).

Our work differs from previous approaches to semantic image compression that use Large Multimodal Models (LMMs)~\cite{MISC, LMM-ITC}. These methods adopt a generative decoder conditioned on both semantic text descriptions and reference images, requiring image codec fine-tuning to enhance compression performance. In contrast, we construct image codecs using the pre-trained AR-based VFMs without additional fine-tuning. Through a comprehensive study of recent VFMs, we make several pioneering contributions: (1) we systematically evaluate the image compression efficiency for a range of VFMs, (2) we provide a detailed analysis to identify the key components within these models that are most critical to compression performance, and (3) we offer novel insights into learned image codecs through the lens of image generation.

%


\section{Related Works}
\subsection{Vision Foundation Models (VFMs)}
Vision foundation models (VFMs) trained with large-scale datasets have shown remarkable performance in generating diverse images with high perceptual realism. Among recent advances in VFMs, AR-based VFMs have emerged as a powerful paradigm. This line of research has culminated in world foundation models, which are capable of generating and simulating complex, dynamic scenes, while implicitly learning highly compressed representations of our visual world~\cite{VJEPA, Cosmos}. The development of AR-based VFMs typically centers on two key components: (1) a visual tokenizer that compresses image into tokens~\cite{ GigaTok, LlamaGen}, and (2) an AR Transformer that models the joint distribution of these tokens. The visual tokenizer converts image to tokens by capturing spatial correlations across image regions to reduce spatial redundancy. The process involves vector quantizing feature vectors into discrete tokens, as exemplified by methods such as VQ-VAE~\cite{VQVAE} and VQ-GAN~\cite{VQGAN}. The AR Transformer then captures token dependencies using two main strategies: next-token and next-scale prediction. The former predicts tokens sequentially in a 1D order and is widely used in VFMs~\cite{LlamaGen, Cosmos, Lumina_mGPT}; Next-scale prediction approach~\cite{VAR} improves the efficiency of next-token prediction by coarse-to-fine predicting strategy: tokens at each scale are generated in parallel, then serve as conditional inputs for predicting tokens at the finer scale.We summarize the key characteristics of these representative models in Table~\ref{tab:VFM}.

\subsection{Learned Image Codecs}
Image compression has evolved from traditional standards like VVC \cite{overview_vvc} to end-to-end optimized neural codecs \cite{googleiclr18, he2022elic}. While these learned methods show promising results, their outputs lack realism, particularly at low bitrates. To bridge this gap, a line of research in generative compression has emerged. These methods prioritize perceptual quality, using adversarial learning \cite{MS-ILLM, HiFiC} or diffusion models \cite{CDC} to produce visually pleasing decoded images. Recent trend involves utilizing large pre-trained models, such as LLMs or Large Multimodal Models, to guide the compression process with semantic understanding. However, these approaches often require designing specialized frameworks or fine-tuning protocols \cite{LMM-ITC, MISC}.
In contrast, our work investigates the intrinsic compression performance of pre-trained, off-the-shelf AR-based VFMs. We show that without any modifications, the powerful predictive mechanisms these models learn for generation can be effective perceptual compressors, revealing a fundamental connection between efficient image generation and compression.

\section{Generation meets Compression}
\subsection{AR-based VFMs as Image Codecs}

\begin{figure}[t]
    \centering
    \includegraphics[width=0.6\linewidth]{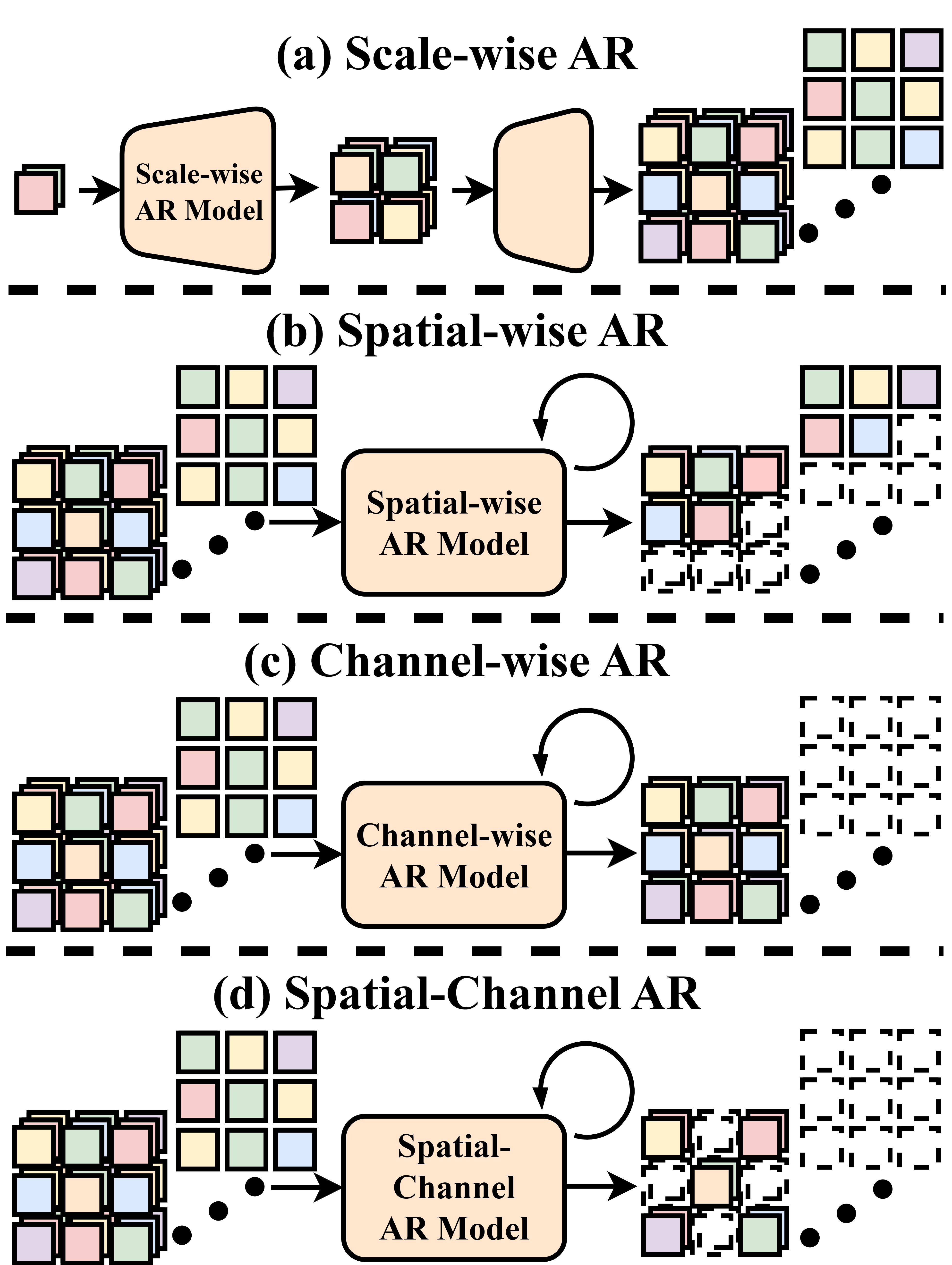}
    \caption{Different types of AR models in learned image codecs.}
    \label{fig:entropy_models}
    \vspace{-1.5em}
\end{figure}

\begin{table*}[t]
\caption{AR-based Vision Foundation Models.}
\centering
\begin{tabular}{l|cccccc}

\hline

\textbf{Models}                & \textbf{Tokenizer} & \textbf{AR model}     & \textbf{Condition} & \makecell{\textbf{\#params} \textbf{(Tokenizer)}}& \makecell{\textbf{\#params} \textbf{(AR)}} & \textbf{Codebook size} \\ \hline
VAR~\cite{VAR}                 & VQ-VAE             & Next-scale prediction & Class              & 108M    & 2.3B                        & 4096                  \\
LlamaGen~\cite{LlamaGen}       & VQ-GAN             & Next-token prediction & Class              & 72M     & 3.1B                        & 16384                 \\
Cosmos~\cite{Cosmos}           & Cosmos tokenizer   & Next-token prediction & Image              & 110M    & 12B                         & 64000                 \\
Lumina-mGPT~\cite{Lumina_mGPT} & VQ-GAN             & Next-token prediction & Image              & 68.7M   & 7B                          & 8192                  \\ \hline
\end{tabular}
\label{tab:VFM}
\vspace{-1em}
\end{table*}



In most VFMs, the token sequences serve a similar function to main image latents in conventional VAE-based image codecs. Moreover, the next-token prediction parallels the conventional spatial context modeling in early learned image codecs~\cite{googleiclr18, he2022elic}, whereas the next-scale prediction (e.g.~in VAR~\cite{VAR}) resembles the hyperprior framework~\cite{googleiclr18}, where a low-scale hyperprior is decoded first and subsequently used to predict the distribution of the higher-scale image latents. It is worth noting that the original hyperprior framework~\cite{googleiclr18} involves only two latent scales, whereas VAR~\cite{VAR} introduces a multi-scale hierarchy. This multi-scale extension has been adopted by the intra-image codec in DHVC~\cite{dhvc}. 



Building on these analogies, we adapt AR-based VFMs for image compression. As shown in Figure~\ref{fig:teaser}(d), the AR model is used to causally predict the coding probability of the next token. The bitrate of each token is estimated based on its self-information. The total rate for a sequence of tokens is then computed as the sum of the individual token rates. Notably, most VFMs exhibit large codebook sizes, which need special considerations when designing their entropy coders, such as arithmetic coders. Most VFMs evaluated in this work were originally developed for conditional generation. To repurpose them for unconditional generation to suit the needs of image compression, we apply model-specific modifications. For image-conditioned models (e.g., Cosmos~\cite{Cosmos} and Lumina-mGPT~\cite{Lumina_mGPT}), we replace the prompt or condition tokens with zero tokens. For class-conditioned models (e.g., VAR~\cite{VAR} and LlamaGen~\cite{LlamaGen}), we follow their configurations and use their specific class tokens for unconditional generation.


\subsection{Learned Image Codecs as Generators}

Typical learned image codecs comprise a primary autoencoder for encoding the main image latents, along with a hyperprior autoencoder that captures side information to facilitate entropy coding. As shown in Fig.~\ref{fig:teaser}(b), to have the image codec function as an unconditional generator, the hyperprior is first sampled from the factorized prior (FP) distribution. This hyperprior is then utilized to estimate the parameters of a conditional Gaussian distribution for each main latent variable. Finally, the main latent is subsequently sampled and decoded to generate the image. Many learned image codecs incorporate AR models to enhance coding efficiency (Fig.~\ref{fig:entropy_models}). These AR models vary in design, capturing dependencies across scales, spatial locations, and channels in distinct ways. These models are based on different assumptions about the factorization of the joint distribution of the main image latents. Scale-wise AR models estimate next-scale latents in a coarse-to-fine manner but often neglect spatial and channel dependencies. In contrast, spatio-channel AR models capture these dependencies more effectively, at the cost of slower generation and heightened error accumulation.
The choice of AR model plays a critical role in determining image generation quality.


\section{Experimental Results}

\begin{table}[t]
\caption{AR Models for Learned Image Codecs and VFMs.}
\centering
\begin{tabular}{l|ccc}
\hline
\textbf{Method}          & \textbf{AR Modeling} & \textbf{Scale} & \textbf{\# params} \\ \hline
ELIC~\cite{he2022elic}   & Spatial-channel      & 2              & 33M                \\
HiFiC~\cite{HiFiC}       & Scale-wise           & 2              & 181M               \\
DHVC~\cite{dhvc}         & Scale-wise           & 4              & 118M               \\
VAR~\cite{VAR}           & Scale-wise           & 10             & 2.3B               \\
LlamaGen~\cite{LlamaGen} & Spatial              & 1              & 3.1B               \\ \hline
\end{tabular}
\label{tab:VFM}
\vspace{-1.5em}
\end{table}

We evaluate the codecs derived from LlamaGen~\cite{LlamaGen}, Cosmos~\cite{Cosmos}, VAR~\cite{VAR}, and Lumina-mGPT~\cite{Lumina_mGPT} on two widely used datasets: Kodak~\cite{kodak} and  CLIC2020~\cite{clic2020}. All images are center-cropped to a resolution of 512×512. Perceptual quality is evaluated using LPIPS~\cite{lpips}, CLIP-IQA~\cite{clipiqa}, and NIQE~\cite{niqe}, while distortion is measured by PSNR and MS-SSIM in the RGB domain. We use the compression ratio to quantify the data reduction contributed by each component in VFM-based codecs. The overall compression ratio is defined as the uncompressed image size divided by the compressed bitrate. For the tokenizer, it is calculated as the image size relative to the token representation without entropy coding (e.g., Cosmos~\cite{Cosmos} has 1024 tokens × 16 bits = 16,384 bits). For the AR model, it is the ratio of the total token size without entropy coding to that achieved with entropy coding.

For comparison, we include a diverse set of representative image codecs, encompassing both learned and conventional designs. These include ELIC~\cite{he2022elic}, VTM~\cite{vtm}, JPEG AI~\cite{jpegai} as distortion-optimized codecs; MS-ILLM~\cite{MS-ILLM} and PerCo (SD)~\cite{perco_sd} as perceptual quality-optimized codecs; and LMM-ImageTextCoding~\cite{LMM-ITC} as a representative image codec that utilizes large multimodal models (LMM) for image compression. 
\subsection{Rate-Distortion Comparison}
\begin{figure*}[t]
    \centering
    \includegraphics[width=0.92\linewidth, trim=0 0 0 0, clip]{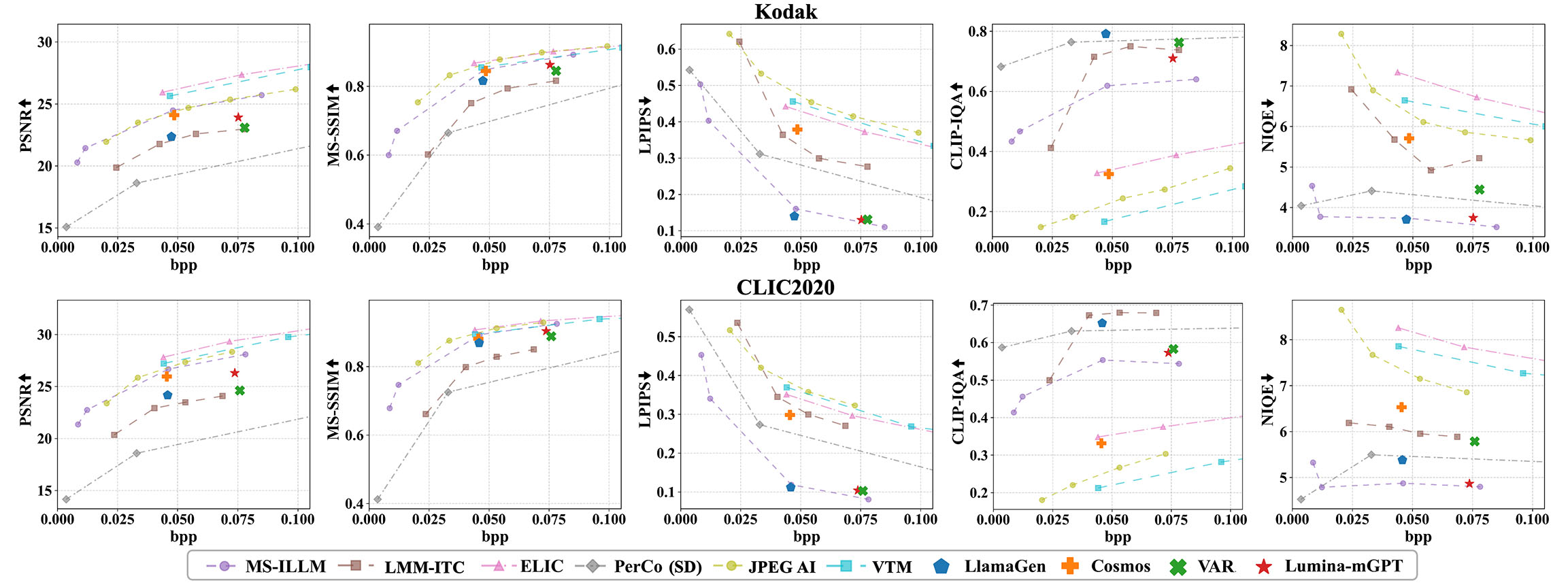}
    \vspace{-0.1cm}
    \caption{Rate-distortion performance comparison on Kodak and CLIC2020.}
    \vspace{-1em}
    \label{fig:rd}
\end{figure*}
\begin{figure*}[t]
    \centering
    \includegraphics[width=0.9\linewidth, trim=0 0 0 0, clip]{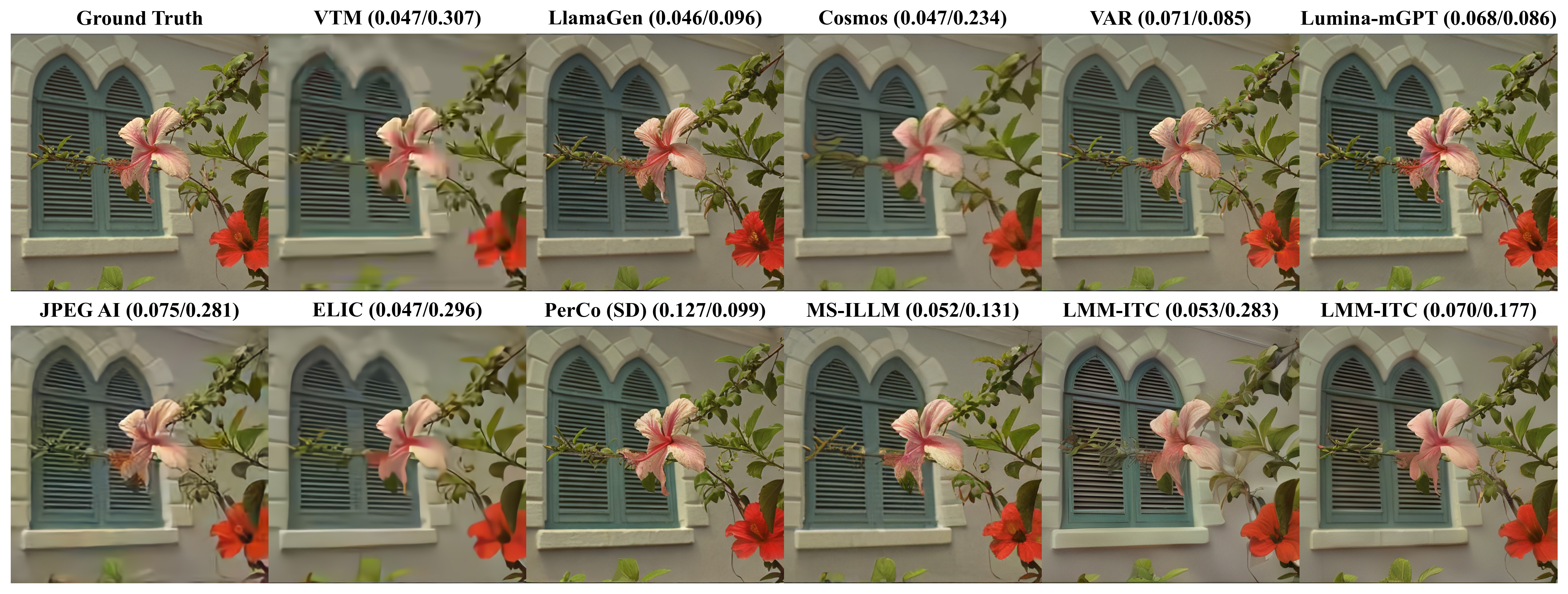}
    \vspace{-0.1cm}
    \caption{Visual comparison of VFMs-codecs and baselines (bpp, LPIPS). More visualization in \href{https://github.com/huutaiphung/VFM-based-codecs}{Github}  .}
    \label{fig:vis}
    \vspace{-1em}
\end{figure*}
\begin{table}[t]
\caption{Coding time for Learned Image Codecs and VFM-codecs.}
\centering
\begin{tabular}{l|ccccc}
\hline
\textbf{Model} & \textbf{ELIC} & \textbf{VAR} & \textbf{LlamaGen} & \textbf{Cosmos} & \textbf{Lumina-mGPT} \\ \hline
Enc. (s)   & 0.13          & 0.23         & 16.79             & 30.32           & 56.11                \\
Dec. (s)   & 0.08          & 0.29         & 16.41             & 30.29           & 97.08               \\ \hline
\end{tabular}
\label{tab:enc_time}
\vspace{-2em}
\end{table}
Fig.~\ref{fig:rd} presents the rate-distortion performance across evaluation metrics. The following observations can be made.
(1) VFM-based codecs demonstrate the ability to achieve exceptionally low bitrates (below 0.1 bpp) on the evaluated test datasets. (2) Without any fine-tuning, most VFM-based codecs surpass distortion-optimized baselines (VTM~\cite{vtm}, JPEG-AI~\cite{jpegai}, ELIC~\cite{he2022elic}) and comparable to perception-optimized codecs (MS-ILLM~\cite{MS-ILLM}, PerCo (SD)~\cite{perco_sd}) in terms of LPIPS. (3) These VFM-based codecs preserve high level of fidelity, as indicated by PSNR and MS-SSIM scores. (4) In terms of no-reference quality metrics, including NIQE and CLIP-IQA, VFM-based codecs exhibit performance comparable to or exceeding that of conventional baselines. (5) When compared to large multimodal model-based codecs (LMM-ITC~\cite{LMM-ITC}), VFM-based codecs achieve superior performance across the majority of assessed metrics. 

Fig.~\ref{fig:vis} presents a visual comparison of the reconstructed images. Most VFM-based codecs can produce high-quality reconstructions with noticeably sharper and clearer structural detail. Notably, LlamaGen achieves better LPIPS scores compared to LMM-ITC~\cite{LMM-ITC} and MS-ILLM~\cite{MS-ILLM} at similar bitrates. Table~???\ref{tab:enc_time} shows the coding time for VFM-based compression compared to learned image codec. The substantial coding times show a complexity bottleneck for VFM-codec that can be furtherly improved. 





\subsection{Ablation study}
\begin{figure}[t]
    \centering
    \includegraphics[width=0.92\linewidth, trim=0 0 0 0, clip]{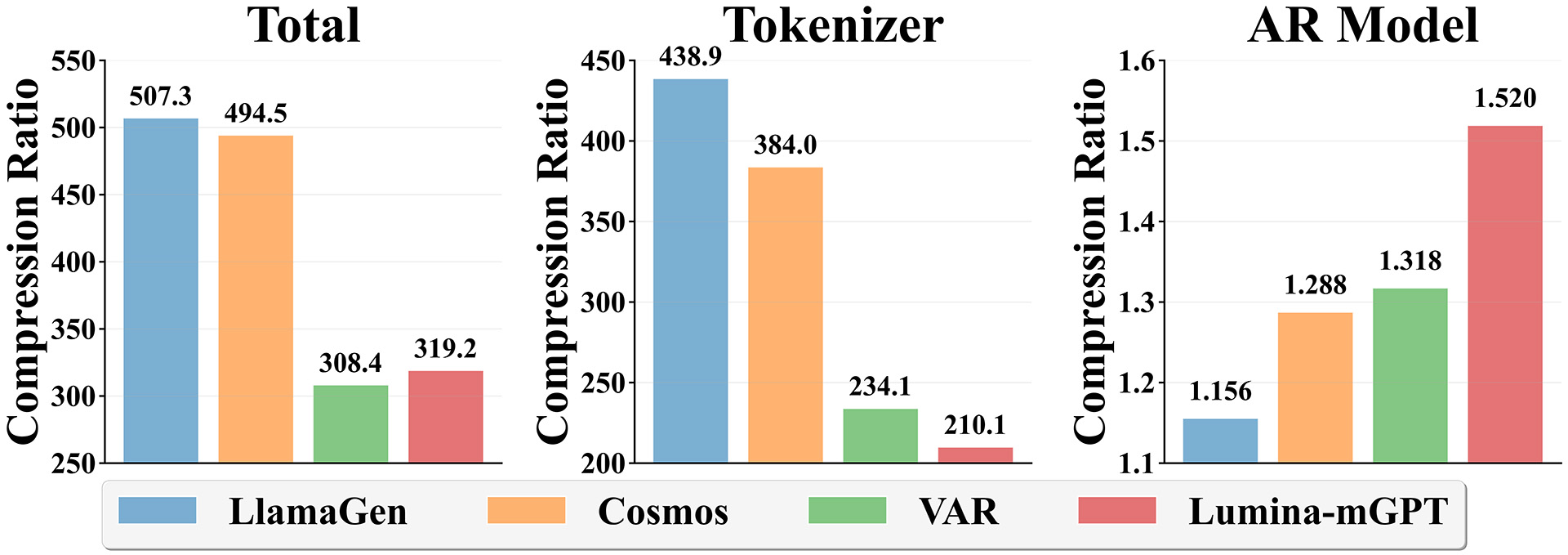}
    \caption{Compression ratios accross VFMs.}
    \label{fig:ablation}
\vspace{-1.5em}
\end{figure}

Fig.~\ref{fig:ablation} presents the overall compression ratios achieved by VFM-based codecs on the Kodak dataset, along with the respective contributions from the tokenizer and AR model. The VFM-based codecs attain overall compression ratios in the range of 300 to 500. A comparison of compression ratios contributed by the tokenizer and AR model reveals that VFMs with lower tokenizer compression ratios tend to gain more from their AR models. We note that the compression ratio of the AR model is influenced by its codebook size (Table~\ref{tab:VFM}). Typically, larger codebooks require more extensive training data to mitigate context dilution in context/AR-based entropy coding. Consequently, smaller codebooks are more favorable for entropy coding efficiency. This trend may explain the AR compression ratios observed in Fig.~\ref{fig:ablation}.




\subsection{Image Generation with Learned Image Codecs}
\label{sec:}
\begin{figure}[t]
    \centering
    \includegraphics[width=0.8\linewidth, trim=0 0 0 0, clip]{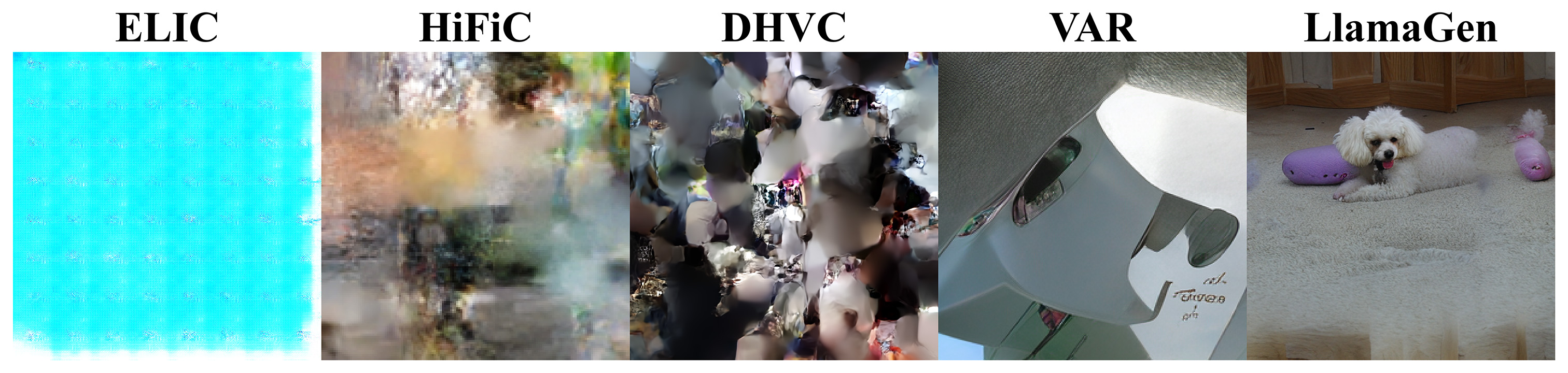}
    \vspace{-0.1cm}
    \caption{Unconditional generation results.}
    \label{fig:vis_generation}
    \vspace{-1.5em}
\end{figure}

\begin{figure}[t]
    \centering
    \includegraphics[width=1\linewidth, trim=0 0 0 0, clip]{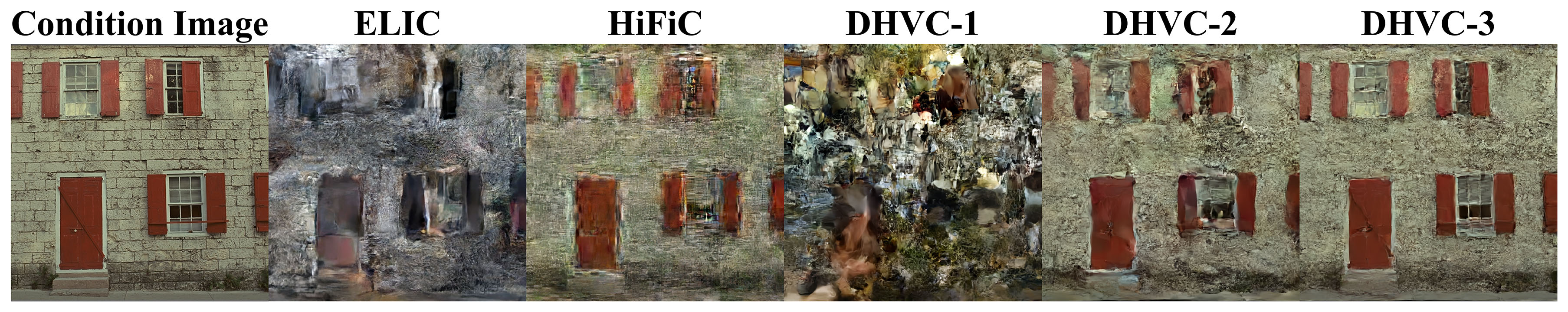}
    \vspace{-0.4cm}
    \caption{Conditional generation results.}
    \label{fig:vis_conditional_generation}
\vspace{-2em}
\end{figure}

Fig.~\ref{fig:vis_generation} presents a comparison of class-conditioned VFMs--VAR~\cite{VAR} and LlamaGen~\cite{LlamaGen}--with learned image codecs in unconditional image generation. To ensure that the comparison is not influenced by low-quality decoders, we use the high-rate settings of HiFiC~\cite{HiFiC} and DHVC~\cite{dhvc}. Table~\ref{tab:VFM} summarizes the characteristics of AR models in these methods. Two key aspects of AR modeling that impact image generation quality are spatial dependencies and channel dependencies. We observe that spatial dependencies are poorly preserved in HiFiC~\cite{HiFiC}, DHVC~\cite{dhvc}, and VAR~\cite{VAR}, leading to unstructured image outputs with diminished semantic consistency. This is due to hyperprior latents being independently sampled from a factorized distribution in learned image codecs, which fails to maintain spatial dependencies. A similar issue is presented in VAR~\cite{VAR}. In contrast, LlamaGen~\cite{LlamaGen} produces more coherent results because it adopts spatial-wise next-token prediction, effectively modeling dependencies between tokens along the spatial dimension and generating more structured and visually coherent images.

We also note that HiFiC~\cite{HiFiC} and DHVC~\cite{dhvc} implement independent scalar quantization of feature samples along the channel dimension, in contrast to vector quantization used in VFMs. Scalar quantization fails to model channel dependencies, often leading to noisier generated images. In comparison, VFMs (e.g.~VAR~\cite{VAR} and LlamaGen~\cite{LlamaGen}) adopt vector quantization, which better preserves channel correlations within each codeword and produces cleaner outputs in local regions. Although ELIC~\cite{he2022elic} adopts a spatial-channel AR model, it still struggles to generate coherent outputs. As shown in~\cite{he2022elic}, ELIC exhibits an energy compaction property along the channel dimension. When early channels are not predicted properly, typically due to sampling from a noisy hyperprior, the resulting errors may propagate catastrophically to subsequent channels.

To validate that sampling hyperpriors from a factorized distribution can degrade image structure, we use hyperpriors encoded from a clean image to estimate the main latents' distributions. In Fig.~\ref{fig:vis_conditional_generation}, HiFiC~\cite{HiFiC} generates a more structured image, although it remains noisy due to its factorial modeling of the main latents across both spatial and channel dimensions. DHVC~\cite{dhvc}, which incorporates three multi-scale hyperpriors, produces increasingly structured results as its hyperpriors are progressively initialized from the input image. ELIC~\cite{he2022elic} also benefits similarly from a more structured hyperprior.  




\section{Conclusion}
This work presents the first study on the compression capabilities of VFMs and reveals the limitations of existing image codecs in image generation. Our findings are as follows: (1) pre-trained VFMs can achieve compression performance comparable to or even better than state-of-the-art image codecs at low bit rates, although their complexity limitations still need to be considered; (2) the tokenizers in VFMs contribute predominantly to the overall compression ratio; (3) conventional image codecs struggle to generate structurally meaningful images due to factorized hyperprior sampling and scalar quantization. These findings open up a new avenue to connect VFMs and learned image codecs for low-rate semantic image compression.



\bibliographystyle{IEEEtran}
\bibliography{IEEEabrv,paper.bib}
\end{document}